# Hexavalent (*Me* - W/Mo)-modified (Ba,Ca)TiO$_3$-Bi(Mg,*Me*)O$_3$ perovskites for high-temperature dielectrics

Thomas Schulz[1], Vignaswaran K. Veerapandiyan[2], Theresa Gindel[2], Marco Deluca[2], Jörg Töpfer[1]

1 – Ernst-Abbe-Hochschule Jena, Dept. SciTec, C.-Zeiss-Promenade 2, 07745 Jena, Germany

2 – Materials Center Leoben Forschung GmbH, Roseggerstrasse 12, Leoben 8700, Austria

Corresponding author:

    Jörg Töpfer

    Univ. Appl. Sciences Jena

    Carl-Zeiss-Promenade 2, 07745 JENA, GERMANY

    Tel.:  +49-3641-205479  Fax.:  +49-3641-205451

    e-mail:    joerg.toepfer@eah-jena.de

**Abstract**

We report on the synthesis of complex lead-free perovskite-type (1-x)(Ba$_{0.8}$Ca$_{0.2}$)TiO$_3$-xBi(Mg$_{0.75}$W$_{0.25}$)O$_3$ (BCT-xBMW) and (1-x)(Ba$_{0.8}$Ca$_{0.2}$)TiO$_3$-xBi(Mg$_{0.75}$Mo$_{0.25}$)O$_3$ (BCT-xBMM) solid solutions via conventional solid-state reaction route. The sintering temperature was adjusted as a function of composition *x* to obtain dense samples (relative densities over 95%) at the same time minimizing bismuth evaporation. X-ray diffraction analysis shows formation of single-phase perovskites for 0 ≤ *x* ≤ 0.10 in the BCT-xBMW series and increasing concentrations of impurity phases for *x* ≥ 0.15 and for *x* ≥ 0.05 in BCT-xBMM. A transition from a tetragonal to pseudo-cubic perovskite structure is observed in BCT-xBMW and BCT-xBMM at *x* = 0.05. The dielectric response has been characterized between -60 °C and 300 °C for BCT-xBMW, and between 30 °C and 300 °C for BCT-xBMM using impedance spectroscopy, showing a transition from ferroelectric to relaxor-like behavior at *x* ≥ 0.05. Additional polarization and Raman spectroscopy measurements reveal the occurrence of highly disordered systems. Analysis of the Raman spectra indicates structural phase changes and lattice modifications caused by chemical substitution. For the composition 0.8Ba$_{0.8}$Ca$_{0.2}$TiO$_3$-0.2Bi(Mg$_{0.75}$W$_{0.25}$)O$_3$ a temperature-stable permittivity of about 600 (±15 % between -60 °C and 300 °C) and small losses of tan$\delta$ < 0.02 for *T* ≤ 230 °C at 1 kHz are observed, making it a suitable dielectric material for high temperature capacitors.

**1. Introduction**

The need for integration of electronics in technical and chemical process monitoring requires new electronic components able to operate in harsh environments, including high temperatures. Multilayer ceramic capacitors (MLCC) are among the most important passive components used in all electronic circuits. Environment-friendly and commercially available MLCCs are limited to working temperatures of up to 150 – 200 °C (X8R, X9R)[1,2]. Different industries, e.g. avionics, automotive, power electronics and downhole or geothermal, are demanding new capacitor components with higher operating temperatures[3,4]. Dielectric materials with stable dielectric properties, i.e. high and temperature-stable dielectric permittivity from -55 °C to above 200 °C paired with low dielectric losses (tan$\delta$ < 0.02), high resistivity and electric-field breakdown strength are thus needed[5].

Classic $BaTiO_3$ (BT) -based ferroelectric materials exhibit high relative permittivity $\varepsilon_r$, but have a strong temperature dependence of $\varepsilon_r$ close to the crystallographic phase transitions and a rapid decay of $\varepsilon_r$ above the Curie point ($T_c$)[6]. One approach to address this problem is to use BT-based solid solution materials, which form relaxor dielectrics[7–11]. In contrast to ordinary ferroelectrics with a sharp transition from the ferroelectric to the paraelectric phase, relaxor materials show broad $\varepsilon_r$ peaks over a wide temperature range with a maximum of the permittivity at temperature $T_m$. The relative permittivity $\varepsilon_r$ of canonical relaxors shows a frequency dispersion at temperatures below $T_m$, while $\varepsilon_r(T)$ exhibit frequency-independent characteristics at $T_m < T < T_B$ ($T_B$ - Burns temperature). At temperatures of T > $T_B$ the permittivity $\varepsilon_r$ of relaxors exhibits Curie-Weiss-behavior like paraelectric dielectrics[12]. Although the exact nature of the relaxor behavior is still in discussion today, the theory of the formation of polar nano regions (PNR) upon cooling below $T_B$ is widely accepted. PNR are described as localized, nano-scaled clusters of polarizations caused from compositional disorder in the crystal lattice[10,11].

Furthermore, BT-based solid solutions with bismuth-containing Bi(*Me*)O$_3$ perovskites have gained special attention as some of them show suppressed relaxor characteristics at high temperatures well above 200 °C, resulting in a flat $\varepsilon_r(T)$ response, making them well suited for high temperature applications[1,2,13,14]. In such systems it is possible either to substitute A-site and/or B-site of the perovskite structure to alter the dielectric response of these materials. The transition from ferroelectric to a relaxor state mainly depends on type (difference in size and valency) and molar fraction of the substituents[15–17]. Substituents with a larger difference in valency compared to the parent lattice ions have a stronger impact on the transition point than smaller or larger ions[11]. Tinberg et al.[18] and Ogihara et al.[19,20] were among the first to report on novel BaTiO$_3$ – Bi(*Me*$^{3+}$)O$_3$ systems modified with trivalent Sc$^{3+}$ ions on B-sites forming a (1-x)BaTiO$_3$ – xBiScO$_3$ solid solution. One of the interesting features of these materials (with x = 0.2 – 0.3) is that they exhibit a weak temperature dependence of permittivity over a wide temperature range with low dielectric losses[19,20]. These compositions are classified as weakly coupled relaxors, because - in contrast to classic relaxors - isolated polar nano clusters show weak coupling only with neighboring clusters at low temperatures and high field conditions[19]. Recently, substitution of trivalent cations (*Me*$^{3+}$) such as Y$^{3+}$ [21,22], Al$^{3+}$ [23] at the B-site of Bi(*Me*$^{3+}$)O$_3$ in BaTiO$_3$ – Bi(*Me*$^{3+}$)O$_3$ system have been investigated. Mixtures of heterovalent cations Bi(*Me'Me''*)O$_3$ instead of isovalent substituents on B-site turned out to be beneficial for engineering the dielectric response at high temperatures. Bivalent (*Me*$^{2+}$) - tetravalent (*Me*$^{4+}$) co-substituents were studied forming BaTiO$_3$ – Bi(*Me*$^{2+}_{1/2}$*Me*$^{4+}_{1/2}$)O$_3$ solid solutions. Combinations of Mg$^{2+}$ [24,25] or Zn$^{2+}$ [26,27] bivalent substituents with Ti$^{4+}$ tetravalent cations as well as with Zr$^{4+}$ [28,29] and recently with Sn$^{4+}$ [30,31] have also been studied. Co-substitution of pentavalent cations like $Nb^{5+}$ [32–34] or $Ta^{5+}$ [35,36] for $Me^{5+}$ in $BaTiO_3$ – $Bi(Mg_{2/3}Me^{5+}_{1/3})O_3$ were also shown to give stable dielectric characteristics over a wide temperature range. Combinations with hexavalent ($Me^{6+}$) substituents, on the other hand, have hardly been reported. Chen et al. reported initial results on the series $BaTiO_3$ – $Bi(Mg_{0.75}W_{0.25})O_3$ and $BaTiO_3$ – $Bi(Zn_{0.75}W_{0.25})O_3$ with $0 \leq x \leq 0.2$ [37,38].

Furthermore, additional incorporation of $Ca^{2+}$-cations on the A-site of BT seems to be beneficial for achieving temperature-stable dielectric characteristics at elevated temperatures. The addition of calcium in the parent BT lattice leads to a shift of the tetragonal-orthorhombic and orthorhombic-rhombohedral phase transitions toward lower temperatures thus stabilizing the tetragonal phase of BT. The cubic-tetragonal transition at $T_C$ (≈ 125 °C) is nearly unaffected although the dielectric peak is less pronounced[14,39,40]. Previously, it was shown that Ca-modified $BaTiO_3$ – $Bi(Mg_{0.5}Ti_{0.5})O_3$ solid solutions[41] exhibit an extended operating temperature range with a small temperature variation of permittivity and a lower usable temperature limit compared to the unmodified system[24,25].

Here we present results on the substitution series $Ba_{0.8}Ca_{0.2}TiO_3$ – $Bi(Mg_{0.75}Me^{6+}_{0.25})O_3$ with $Me^{6+}$= W and Mo. This work aims to expand the understanding on how bismuth perovskites with mixed bivalent $Mg^{2+}$- hexavalent - $Mo^{6+}$ or $W^{6+}$ cation substitutions affect the dielectric and structural properties of $Ba_{0.8}Ca_{0.2}TiO_3$. The material synthesis, phase analysis, dielectric properties and structural properties are reported here for these complex perovskite oxides.

## 2. Experimental Procedure

Bulk ceramic samples of (1-x)Ba$_{0.8}$Ca$_{0.2}$TiO$_3$-xBi(Mg$_{0.75}$W$_{0.25}$)O$_3$ (BCT-xBMW) and (1-x)Ba$_{0.8}$Ca$_{0.2}$TiO$_3$-xBi(Mg$_{0.75}$Mo$_{0.25}$)O$_3$ (BCT-xBMM) with 0 ≤ *x* ≤ 0.25 were prepared using the conventional solid-state reaction route. Starting materials were reagent grade barium carbonate (BaCO$_3$, Solvay HP BM040, ≥ 99.0 %), calcium carbonate (CaCO$_3$, Merck EMSURE®, > 98.5 %), titanium dioxide (TiO$_2$, Evonik Aeroxide P90, ≥ 99.5 %), bismuth(III) oxide (Bi$_2$O$_3$, Sigma Aldrich, 99.9 %), magnesium oxide (MgO, Merck EMSURE®, ≥ 97 %), molybdenum(VI) oxide (MoO$_3$, Alfa Aesar, 99.5 %), tungsten(VI) oxide (WO$_3$, Alfa Aesar, 99.8 %, specific surface area ≈ 0.85 m²/g). A nanoscale tungsten(VI) oxide powder (WO$_3$, Aldrich, specific surface area ≈ 7.4 m²/g) was tried as an alternative raw material for W$^{6+}$ source to assist the phase formation process. CaCO$_3$ was calcined at 900 °C for 4 hours to form CaO before weighing in. Bi$_2$O$_3$ was dried at 700 °C for 4 hours to decompose Bi$_2$O$_2$CO$_3$ impurities. A stoichiometric ratio of the raw materials of all compositions were weighed and homogenized (30 g batches) in isopropanol for 12 hours with stabilized zirconia balls (YTZ®, TOSOH) in a roll mill (100 rpm). After homogenization, the slurry was dried in a rotary evaporator followed by calcination of the powders between 900 – 1100 °C for 3 hours in covered alumina crucibles to minimize bismuth losses through volatilization. After calcination the powders were ball milled in isopropanol using a planetary ball mill (Pulverisette 5, Fritsch GmbH) for 2 – 3 hours in polyoxymethylene (POM) grinding beakers with zirconia grinding media (diameter 1 mm). After achieving particle sizes between 0.6 – 0.7 µm, measured using a laser diffraction analyzer (Mastersizer 2000, Malvern Panalytical Ltd.), the slurry was cast through a 180 µm nylon sieve and dried.

The calcined and milled powders were mixed with 1.5 wt% polyvinyl alcohol binder and uniaxially pressed at 37 MPa into disks of 13 mm in diameter and about 2 mm in thickness. Sintering of the samples was performed in a muffle furnace at various temperatures between 1150 °C and 1350 °C depending on the composition for 3 hours with uniform heating and cooling rates of 5 °C/min as summarized in *Table 1* for both systems. To reduce bismuth loss, the pellets were buried in a powder bed of same composition (each pellet of 0.7 g covered with 0.3 g powder) and covered with an alumina crucible.

To study the shrinkage behavior of powder compacts (pellets of 10 mm diameter and 3 mm height, 60 MPa compaction pressure) during sintering, dilatometry was performed using a push-rod dilatometer (DIL 402E, Netzsch GmbH) with a heating rate of 4 °C/min. The density of sintered samples was determined using Archimedes' method. Powder X-ray diffraction measurements (XRD) were performed using a Bruker D8 Advance diffractometer on powders obtained by crushing the sintered pellets (roughly 0.5 g powder on Si low background sample holders). CuKα radiation, 40 kV, 40 mA, stepsize 0.015° and 3 s per step were used to record the diffractograms. Lattice parameters were calculated using the software TOPAS (Bruker Corp.). Microstructural analysis was performed on polished and thermally etched samples using scanning electron microscopy (SEM, Ultra 55, Carl Zeiss Microscopy GmbH).

For dielectric characterization, the pellets were polished down to 1 mm thickness and gold electrodes were applied on both sides via a sputtering process. Impedance Spectroscopy (Dielectric Broadband Spectrometer BDC with Solartron SI1260, Novocontrol Technologies GmbH & Co. KG) was used for measuring the temperature dependent permittivity and loss tangent between -60 °C and 300 °C, using a heated liquid nitrogen stream for measurements below room temperature, and a heated airflow for higher temperatures. Polarization versus electric field measurements (P-E loops) were conducted using a custom-made Sawyer-Tower circuit. The samples were immersed in silicon oil. A HAMEG oscilloscope, TREK amplifier and a triangular voltage signal (function generator DS345, Stanford Research Systems) at a frequency of 13 Hz were used to measure polarization as a function of applied electric field. Raman measurements were carried out in a LabRAM 300 spectrometer (Horiba Jobin Yvon) using an excitation wavelength of 532 nm in a backscattering geometry. Temperature dependent Raman measurements were carried out using a Linkam (THMS600) temperature-controlled stage.

## 3. Results and discussion

The shrinkage behavior was examined using dilatometry in order to determine the optimum sintering conditions for the BCT-xBMW and BCT-xBMM samples. The shrinkage curves of powder compacts exhibit a trend towards decreasing sintering temperatures with increasing $x$ for BCT-xBMW (*Fig. 1*). For unsubstituted $Ba_{0.8}Ca_{0.2}TiO_3$ ($x = 0$) maximum shrinkage appears at about 1280 °C and shrinkage terminates at around 1330 °C. For substituted perovskites with $x > 0$ shrinkage tends to appear at lower temperatures. Samples with $0 \leq x \leq 0.1$ were sintered at 1350 °C in order to achieve dense samples with relative density of $\geq 95$ %. To minimize Bi evaporation, samples with larger substitution concentration were sintered at temperatures lower than 1350 °C, with temperatures shown to be sufficient for producing dense samples (*Table 1*).

Room temperature XRD patterns of the sintered samples of BCT-xBMW with $0 \leq x \leq 0.25$ are shown in *Fig. 2 (a)*. The patterns show a single perovskite phase for compositions with $x \leq 0.10$. For $x \geq 0.15$ low-intensity peaks of a $BaWO_4$ impurity phase were identified. In an attempt to reduce the concentration of the impurity phase, nano-size tungsten (VI) oxide powder ($WO_3$, Aldrich, < 100 nm) was tested as an alternative raw material for $x = 0.2$. However, similar impurities were observed in XRD patterns after sintering, hence standard $WO_3$ was used as tungsten source for all compositions. *Fig. 2 (b)* shows the enlarged XRD patterns in the 2θ-range between 44° and 47° which represents the (002)/(200) lattice planes of the perovskite structure. For BCT-xBMW samples with $x \geq 0.05$ the (002)/(200) peaks merge into one peak indicating a phase transformation from tetragonal into a pseudo-cubic perovskite structure. *Fig. 2 (c)* and *(d)* show the diffraction patterns for the BCT-xBMM sample series. In contrast to the BCT-xBMW system, the $BaMoO_4$ impurity phase was detected at a lower concentration of $x \geq 0.05$ in addition to peaks of the perovskite main phase. At larger xBMM concentrations of x = 0.20 and x = 0.25 low-intensity peaks of other impurity phases (bismuth-rich phase $Bi_{8.11}Ba_{0.89}O_{13.05}$ (PDF 45-0289), $CaMoO_4$ (PDF 85-1267) and $CaTiO_3$ (PDF 78-1013)) were detected. Variations of the calcination and sintering temperatures did not improve the phase purity of the BCT-xBMM perovskite series. It can be concluded that the formation of a single-phase BCT-xBMM solid solution via solid state route seems to be possible in a very limited compositional range only.

*Fig. 3* shows the variation of the calculated lattice parameters and cell volumes for (a) BCT-xBMW and (b) BCT-xBMM compositions. Lattice parameters were refined from room temperature XRD patterns. Tetragonal (s.g. P4mm) lattice parameters were obtained for BCT-xBMW and BCT-xBMM compositions with $x \leq 0.05$, while for $x > 0.05$ pseudo-cubic (s.g. $Pm\bar{3}m$) parameters were determined. For both systems the tetragonality decreases with increasing substitution and becomes very small at $x \geq 0.05$ showing a transition to a pseudo-cubic structure within the limits of XRD measurements. In BCT-xBMW samples $a_0$ increases with increasing $x$. The constant increase of the $a_0$ indicates the formation of a solid solution in the whole compositional range despite the observation of small concentrations of secondary phase at high xBMW content. In accordance with the trend of $a_0$ the cell volume also increases with increasing xBMW addition. For the BCT-xBMM samples, a similar trend with increasing substitution is shown in *Fig. 3 (b)* until composition $x = 0.2$. For the composition xBMM at $x = 0.25$ the lattice parameter $a_0$ as well as the cell volume seems to reach a limit, which is consistent with the thesis of a limited solubility of xBMM in the parent BCT lattice.

The microstructure and corresponding EDX mappings for Bi, Mg and W of BCT-xBMW samples with $x = 0.1$ and $x = 0.2$ are illustrated in *Fig. 4 (a & b)*. The EDX maps of Ba and Ti are not shown due to peak overlap. The samples have a dense microstructure with a relative density higher than 95 % (cf. *Table 1*). *Fig. 4 (a)* shows a homogenous microstructure of the sample $x = 0.1$ with a nearly homogeneous distribution of the chemical components. For $x = 0.2$ a fine-grained microstructure is observed in the BSE micrograph shown in *Fig. 4 (b)*. The smaller grain size as compared to the sample with $x = 0.1$ is caused by the lower sintering temperature (cf. *Table 1*). EDX mappings show an inhomogeneous distribution of components with magnesium- and tungsten-rich areas. The inhomogeneity is consistent with the tungstate impurity phase observed using XRD in that composition. *Chen et al.*[37] investigated $(1-x)BaTiO_3-xBi(Mg_{0.75}W_{0.25})O_3$ (BT-xBMW) and reported single-phase perovskite formation in the range $0 < x < 0.24$. However, in the BCT-based system studied here, $BaWO_4$ impurities are identified for $x \geq 0.15$. One possible reason for the difficulty of incorporating W-ions in the lattice is due to overall reduction in unit cell volume of the BCT-xBMW system as compared to the BT-xBMW system. The $Ca^{2+}$ -ions ($r_{Ca2+}$ = 1.34 Å) have a smaller ionic radius as compared to $Ba^{2+}$-ion ($r_{Ba2+}$ = 1.61 Å)[42] resulting in smaller pseudo-cubic lattice parameters for the BCT-xBMW system (*Fig. 3 (a)*) compared to those of the BT-xBMW system[37]. Compared with *Fig. 4 (b)* the BSE image and EDX mappings for BCT-xBMM (*x* = 0.2) are shown in *Fig. 4 (c)*. The microstructure of this sample consists of smaller grains which is consistent with the lower sintering temperature ($T_{sint.}$ = 1150 °C) used to achieve a relative density higher than 95 %. In *Fig. 4 (c)*, an inhomogeneous distribution of the substituents is observed with enriched areas of magnesium and molybdenum. Such segregations seem to strongly affect the dielectric characteristics of this material, which is discussed later.

The dielectric properties of BCT-xBMW compositions, as determined using impedance spectroscopy, are plotted in *Fig. 5*. The parent $Ba_{0.8}Ca_{0.2}TiO_3$ (BCT) is consistent with ferroelectric behavior, *Fig. 5 (a)*, as indicated by the dielectric anomaly peak with a Curie-temperature $T_C$ at 125 °C. Contrary to $BaTiO_3$, the compound BCT does not show any dielectric anomaly at around 5 °C, as anticipated since $Ca^{2+}$ substitution strongly shifts the orthorhombic to tetragonal phase transition to lower temperatures[39,43]. Addition of 5 mol% BMW results in a shift of the permittivity peak to lower temperature and broadening of the ferroelectric anomaly. The frequency dispersion of the permittivity maximum ($T_m$) and the frequency-independent part at higher temperatures are typical features of relaxor-like behavior[12]. Further increase of xBMW content results in further broadening of the permittivity response and eventual flattening of the curve with suppressed temperature dependence of the permittivity. Flat permittivity-temperature curves with smaller absolute values of permittivity are observed for compositions with higher amounts of xBMW (*x* = 0.2 and *x* = 0.25).

A summary of the dielectric properties of BCT-xBMW extracted from the impedance spectroscopy measurements are given in *Table 2*. As mentioned above, the compositions with high amounts of xBMW ($x \geq 0.2$) show a small temperature dependence of permittivity ($\leq 15$ %) over a wide temperature range (-60 °C – 300 °C) for a specific frequency, 1 kHz. Dielectric losses are $\leq 0.02$ over the stated temperature range in *Table 2*. However, for large *x* and at high temperatures, increased low-frequency dielectric losses are measured, which are likely linked to the increasing concentration of the impurity phase. Nevertheless, the composition $0.8Ba_{0.8}Ca_{0.2}TiO_3$-$0.2Bi(Mg_{0.75}W_{0.25})O_3$ shows promising results as high temperature capacitor dielectric material with a mid-range permittivity of about 600 and small losses tan$\delta$ < 0.02 from -30 °C to 230 °C (at 1 kHz). As mentioned before, using a nanoscale $WO_3$ as a raw material for the composition 0.8BCT-0.2BMW did not alter the dielectric properties significantly.

For BCT-xBMM compositions, the temperature dependent permittivity and loss are given in *Fig. 6*. With small addition of xBMM ($x = 0.05$) the permittivity peak of BCT, *Fig. 6 (a)*, is broadened and shows relaxor-like characteristics, *Fig. 6 (b)*. The broadening and suppression of the permittivity peak becomes more prominent with further increase of the xBMM fraction ($x \geq 0.1$) as shown in *Fig. 6 (c – d)*. For compositions with high amounts of xBMM substitution ($x \geq 0.2$), a strong low frequency dispersion occurs at high temperatures (*Fig. 6 (e – f)*) which is attributed to the high amount of impurity phases, see *Fig. 2 (c)* and *Fig. 4 (c)*. Overall, due to the high dielectric loss displayed by all measured samples at *T* > 150 °C, BCT-xBMM is not as an attractive high temperature dielectric as BCT-xBMW. For this reason, further discussion of the data focuses on the BCT-xBMW system.

To further analyze the dielectric response, the Curie-Weiss plot, i.e. the reciprocal permittivity $1/\varepsilon'_r$ at 100 kHz vs. temperature is shown in *Fig. 7 (a)* for all BCT-xBMW samples. Points marked with symbols represent the temperature of maximum permittivity ($T_m$ - blue) and Burns temperature ($T_B$ - red). Values of $T_B$ were determined as temperatures of deviations from the linear Curie-Weiss fit at 100 kHz. For BCT ($x = 0$) which displays a ferroelectric behavior, $T_C = T_m = 125$ °C is observed, whereas for $x > 0$ relaxor-like behavior is found. The variation of $T_m$ and $T_B$ with composition $x$ is depicted in *Fig. 7 (b)*. $T_m$ exhibits a marked drop from $T_C = 125$ °C for $x = 0$ to $T_m = $ -10 °C for $x = 0.05$ at 100 kHz. For larger $x$, $T_m$ of the broad peaks increases slightly. $T_B$ exhibits a constant increase with $x$. For temperatures higher than $T_B$ the reciprocal permittivity follows the Curie-Weiss-law with a linear dependency, which is best described as a paraelectric state of the material. For the temperature range between $T_m$ and $T_B$ a modified Curie-Weiss-law[44] can be applied (eq. 1) with a diffuseness exponent $\gamma$ that can take values between 1 (indicating normal ferroelectric behavior) and 2 (indicating a complete diffuse phase transition):

$$\frac{1}{\varepsilon'_r} = \frac{1}{\varepsilon'_{r,max}} + \frac{(T - T_m)^\gamma}{C} \tag{1}$$

The diffuseness exponents $\gamma$ for BCT-xBMW compositions were obtained from the slopes of the $\ln(1/\varepsilon'_r - 1/\varepsilon'_{r,max})$ vs $\ln(T - T_m)$ plots and are shown in *Fig. 7 (b)*. The values vary between 1.25 and 2 implying that even the pure $Ba_{0.8}Ca_{0.2}TiO_3$ shows traces of a diffuse phase transition which becomes more prominent with increasing $xBi(Mg_{0.75}W_{0.25})O_3$ addition, denoting a highly disordered system.

Additionally, polarization versus electric field measurements of the BCT-xBMW samples at room temperature are plotted in *Fig. 8*. Pure BCT shows a hysteresis loop with a remanent polarization, which is characteristic of ferroelectric materials. Increasing the xBMW content slims the P-E-loop, as shown for compositions $x = 0.05$ and $x = 0.1$, which is ascribed to a transition from ferroelectric to relaxor ferroelectric.

Further increasing the xBMW substitution ($x > 0.1$) results in a nearly linear P-E-response.

Fig. 9 shows the RT Raman spectra of BCT-xBMW samples. At $x = 0$, the spectrum resembles that of pure BT[45]. The simultaneous presence of an interference dip at around 180 cm$^{-1}$, the B$_1$ mode (mode 4), and the modes 5 and 7 is an indication of the ferroelectric nature of the sample. For $x > 0$, significant changes in the Raman spectral signature occurs. One of the aspects to be noted is the clear indication of drastic increase in the translational disorder of the system, evident from spectral broadening. For instance, the A$_1$-type mode 5 has an asymmetric tail (mode 6) in the higher wavenumber region. Mode 6 is present in all polycrystalline barium titanates[46], but its strong evolution with $x$ indicates a substitution-driven increase in the intrinsic lattice disorder. In addition, it is important to note that the presence of impurity phase as recorded by XRD is also evident in the Raman spectra for the $x \geq 0.20$ (and perhaps $x = 0.15$) compositions by the presence of extra peaks (marked with *). The presence of high wavenumber modes above ~715 cm$^{-1}$ is related to the vibration of oxygen octahedra (BO$_6$) in which atomic species different from Ti are present[47]. The appearance of two separate modes (mode 8 and 9) above mode 7 is thus likely due to the substitution of two different species (Mg and W) at the B-site in the BCT-xBMW material. These modes are broad because of the severe heterogeneity at the B-site in the BCT-xBMW samples. This is evident also from the disappearance of the interference dip at 180 cm$^{-1}$ for $x > 0$ (indicated as mode 2 in Fig. 9), which is related with loss of correlation of Ti displacements (e.g. hindered by B-site disorder[47]). The lattice distortion is not only attributed here to the B-site heterogeneity but also the presence of the electron lone pair of Bi$^{3+}$, substituting at the A-site, which introduces additional octahedral distortion. Bi$^{3+}$ substitution in fact changes the bonding nature of the A-O bond[48] and this influences also the Ti-O bond length, as signaled by the shift of mode 3[49] (which is associated to O-Ti-O vibrations) for *x* > 0, as marked in *Fig. 9*. Interestingly, a hump (labelled as 'a') resembling mode 3 in pure BCT is observed only in x = 0.05 and this is speculated to be a competing effect of $Ca^{2+}$ (homovalent) and $Bi^{3+}$ (heterovalent) at the A-site on the lattice strains resulting from their ability to go off-center. The effect of $Bi^{3+}$ incorporation in the A-site is also observed by the increasing intensity of mode 1 for x > 0. This mode is the result of a two-mode behavior and was previously related to nanosized Bi cation clustering[50].

*Fig. 10* shows the temperature dependent Raman spectra for *x* = 0, 0.05 and 0.25, in order to track down the evolution of phase transitions. Measurements were carried out with special care to avoid any impurity phase at all measured points. As a rule of thumb, the number of allowed Raman modes is dictated by group theory based on the crystal symmetry. The number of modes increases with the decrease in the crystal symmetry, and the disappearance of Raman modes with temperature change is usually an indication of a structural phase transition. This is evident in BCT-xBMW for *x* = 0 by the disappearance of the $B_1$ mode at ~300 $cm^{-1}$ above 130 °C (marked with an arrow in *Fig. 10 (a)*), which is in correspondence with the $T_C$ from the relative permittivity measurements. The presence of a broad spectral signature in the cubic phase above the Curie temperature for *x* = 0 occurs because of thermally activated dynamic disorder in the system[47]. For *x* > 0, at all measured temperatures, the Raman spectrum always looks alike, hence there is no abrupt change of Raman modes at any measured temperatures. This is common in disordered systems that show pseudo-cubic structure at a larger length scale (as measured by laboratory XRD), but are disordered at a shorter length scale (as probed by Raman). The absence of any change in the Raman spectra for *x* > 0 over the measured temperature range suggests that the material is in a relaxor state already at *x* = 0.05 [47]. The Raman data are consistent with the results of dielectric measurements, and thus justify the large jump in diffuseness parameter displayed already at $x = 0.05$ (cf. *Fig. 7 (b)*).

## 4. Conclusion

Limited solid solutions of BCT-xBMW and BCT-xBMM were prepared via the mixed oxide route. The sintering temperatures were adjusted according to concentration $x$ and dense samples ($\rho_{rel.} > 95\ \%$) were obtained. X-ray diffraction measurements reveal the formation of perovskite solid solutions and the appearance of small concentrations of tungstate or molybdate impurity phases at $x_{BMW} \geq 0.15$ and $x_{BMM} \geq 0.05$, respectively. Impedance data, polarization measurements and Raman studies for BCT-xBMW show a transition from a ferroelectric phase to a highly disordered relaxor state with increasing substituent addition. The transition to relaxor takes places at low concentrations of xBMW ($x \leq 0.05$), which is expected in such complex perovskite systems with different ion sizes and large differences in valency state of A- and B-site ions. The dielectric properties of the BCT-xBMW system show promising results as high temperature capacitor dielectric material. The composition 0.8BCT-0.2BMW exhibits a temperature-stable mid-range permittivity of about 600 ($\pm 15\ \%$ between -60 °C and 300 °C) and small losses $\tan\delta < 0.02$ for $T < 230$ °C at 1 kHz.

## Acknowledgements

The work was financially supported within the project "KERBESEN" by the Thuringian Ministry for Economy, Science and Digital Society – European Social Fund Thuringia, Germany (2015FGR0084). V. K. V. and M. D. acknowledge support from the Austrian Science Fund (FWF): Project P29563-N36.

**Figure captions**

Fig. 1: Shrinkage behavior of BCT-xBMW determined using dilatometry indicates reduced starting temperatures of the sintering process with increasing bismuth content

Fig. 2: X-ray powder diffraction patterns and expanded view of 002/200 peaks of BCT-xBMW (a & b) and BCT-xBMM (c & d) ceramics, respectively (impurity phases denoted with stars are discussed in the text)

Fig. 3: Lattice parameters of BCT-xBMW (a) and BCT-xBMM samples (b)

Fig. 4: SEM images and EDX mappings of the elements Bi, Mg, W and Mo for the samples (a) $0.9Ba_{0.8}Ca_{0.2}TiO_3-0.1Bi(Mg_{0.75}W_{0.25})O_3$; (b) $0.8Ba_{0.8}Ca_{0.2}TiO_3-0.2Bi(Mg_{0.75}W_{0.25})O_3$ and (c) $0.8Ba_{0.8}Ca_{0.2}TiO_3-0.2Bi(Mg_{0.75}Mo_{0.25})O_3$

Fig. 5: Temperature-dependent permittivity (solid lines) and loss factors (dashed lines) of BCT-xBMW samples at different frequencies (100 Hz – 1 MHz): (a) x = 0; (b) x = 0.05; (c) x = 0.1; (d) x = 0.15; (e) x = 0.2; (f) x = 0.25

Fig. 6: Temperature-dependent permittivity (solid lines) and loss factors (dashed lines) of BCT-xBMM samples at different frequencies (100 Hz – 1 MHz): (a) x = 0; (b) x = 0.05; (c) x = 0.1; (d) x = 0.15; (e) x = 0.2; (f) x = 0.25

Fig. 7: Curie-Weiss-plot with marked temperature of maximum permittivity ($T_m$) and Burns temperature ($T_B$) (a); trend of $T_m$, $T_B$ and diffuseness exponent γ over compositional variation for BCT-xBMW samples (b) determined from impedance data at 100 kHz

Fig. 8: Polarization vs. field of BCT-xBMW samples at room temperature

Fig. 9: Room temperature Raman spectra of BCT-xBMW ceramics. The peak marked with an asterisk for the x = 0.25 composition denotes the secondary phase

Fig. 10: Temperature dependent Raman spectra for BCT-xBMW: (a) x = 0; (b) x = 0.05 and (c) x = 0.25

**Table captions**

Table 1: Summary of synthesis parameters and densities of BCT xBMW and BCT xBMM samples after sintering

Table 2: Summary of dielectric properties of BCT-xBMW samples measured at 1 kHz

*Table 1: Summary of synthesis parameters and densities of BCT-xBMW and BCT-xBMM samples after sintering*

| Sample | Calcination temperature (°C) | Sintering temperature (°C) | Density (g/cm³) | Relative density (%) |
|---|---|---|---|---|
| $x = 0.00$ | 1100 | 1350 | 5.43 (±0.05) | 96.2 (±0.9) |
| (1-$x$)Ba$_{0.8}$Ca$_{0.2}$TiO$_3$-$x$Bi(Mg$_{0.75}$W$_{0.25}$)O$_3$ | | | | |
| $x = 0.05$ | 1000 | 1350 | 5.62 (±0.02) | 97.5 (±0.3) |
| $x = 0.10$ | 1000 | 1350 | 5.64 (±0.01) | 95.8 (±0.2) |
| $x = 0.15$ | 1000 | 1300 | 5.83 (±0.01) | 96.9 (±0.1) |
| $x = 0.20$ | 1000 | 1300 | 5.94 (±0.01) | 96.7 (±0.2) |
| $x = 0.25$ | 1000 | 1200 | 6.07 (±0.02) | 96.9 (±0.4) |
| (1-$x$)Ba$_{0.8}$Ca$_{0.2}$TiO$_3$-$x$Bi(Mg$_{0.75}$Mo$_{0.25}$)O$_3$ | | | | |
| $x = 0.05$ | 900 | 1350 | 5.33 (±0.02) | 92.7 (±0.3) |
| $x = 0.10$ | 900 | 1300 | 5.84 (±0.02) | 95.7 (±0.8) |
| $x = 0.15$ | 900 | 1200 | 5.76 (±0.02) | 97.1 (±0.4) |
| $x = 0.20$ | 900 | 1150 | 5.85 (±0.01) | 97.1 (±0.1) |
| $x = 0.25$ | 900 | 1150 | 5.89 (±0.02) | 93.6 (±0.3) |

*Table 2: Summary of dielectric properties of BCT-xBMW samples measured at 1 kHz*

| sample | $\varepsilon'_{r,max}$ | $T_m$ (°C) | $\varepsilon'_{r,mid}$ | T-range $\varepsilon'_{r,mid} \pm 15\%$ (°C) | T-range tan$\delta$ < 0.02 (°C) |
|---|---|---|---|---|---|
| *x* = 0.00 | 8559 | 127 | - | - | - |
| *x* = 0.05 | 4793 | -30 | - | - | -20 - 285 |
| *x* = 0.10 | 1502 | -50 | - | - | -20 - 265 |
| *x* = 0.15 | 842 | -40 | 732 | -60 - 190 | -40 - 240 |
| *x* = 0.20 | 676 | 30 | 598 | -60 - 300 | -30 - 230 |
| *x* = 0.25 | 537 | 125 | 484 | -60 - 300 | -25 - 190 |

# Figures

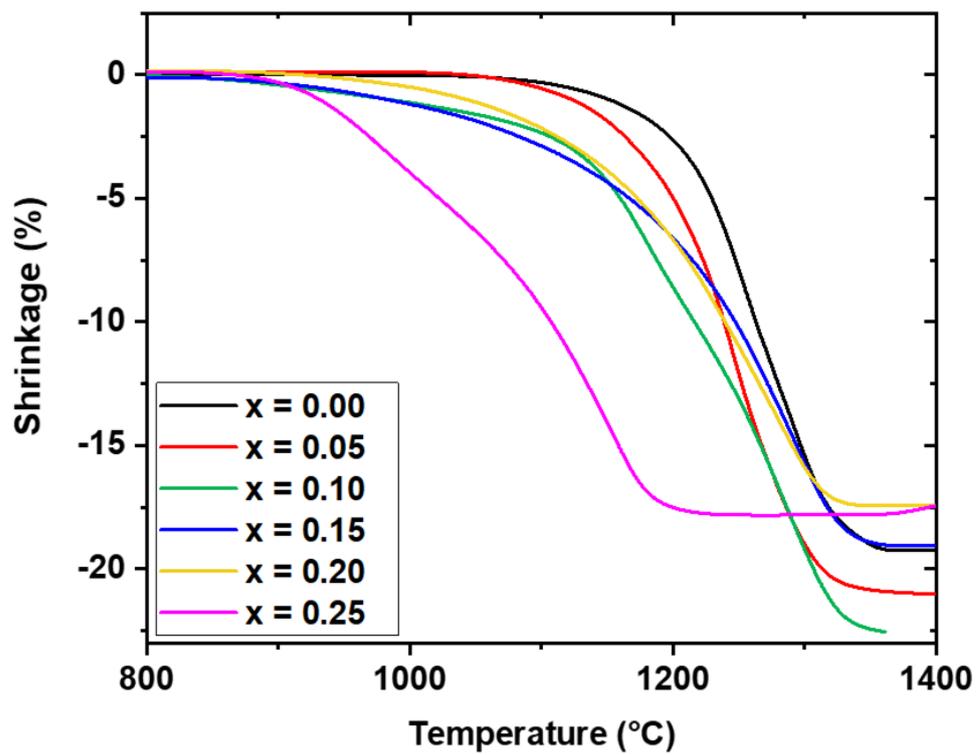

Fig. 1: Shrinkage behavior of BCT-xBMW determined using dilatometry indicates reduced starting temperatures of the sintering process with increasing bismuth content

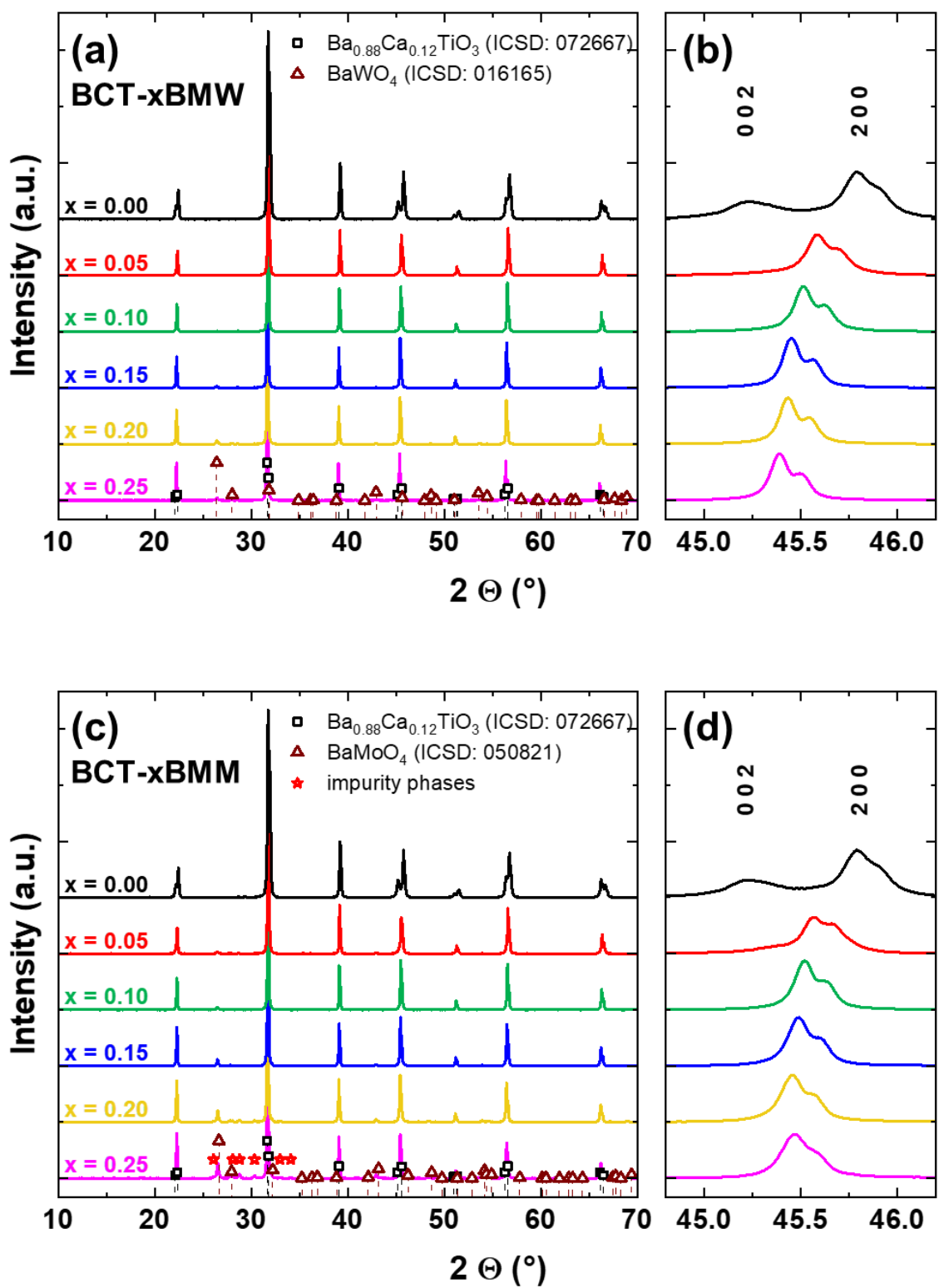

Fig. 2: X-ray powder diffraction patterns and expanded view of 002/200 peaks of BCT-xBMW (a & b) and BCT-xBMM (c & d) ceramics, respectively (impurity phases denoted with stars are discussed in the text)

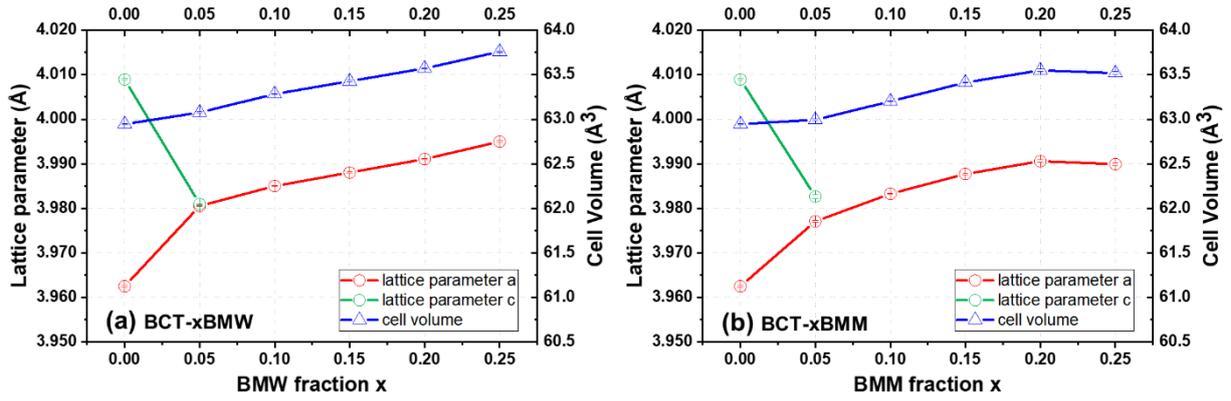

Fig. 3: Lattice parameters of BCT-xBMW (a) and BCT-xBMM samples (b)

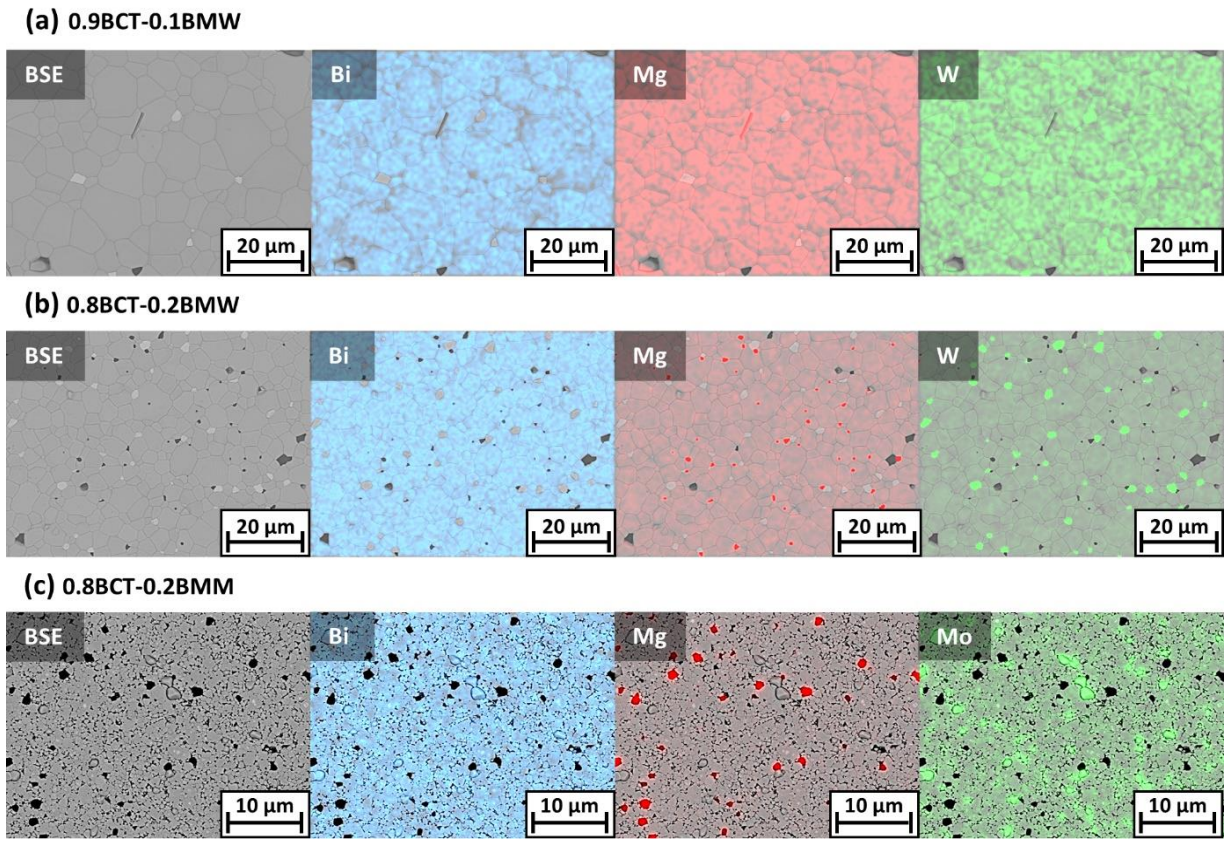

Fig. 4: SEM images and EDX mappings of the elements Bi, Mg, W and Mo for the samples (a) $0.9Ba_{0.8}Ca_{0.2}TiO_3$-$0.1Bi(Mg_{0.75}W_{0.25})O_3$; (b) $0.8Ba_{0.8}Ca_{0.2}TiO_3$-$0.2Bi(Mg_{0.75}W_{0.25})O_3$ and (c) $0.8Ba_{0.8}Ca_{0.2}TiO_3$-$0.2Bi(Mg_{0.75}Mo_{0.25})O_3$

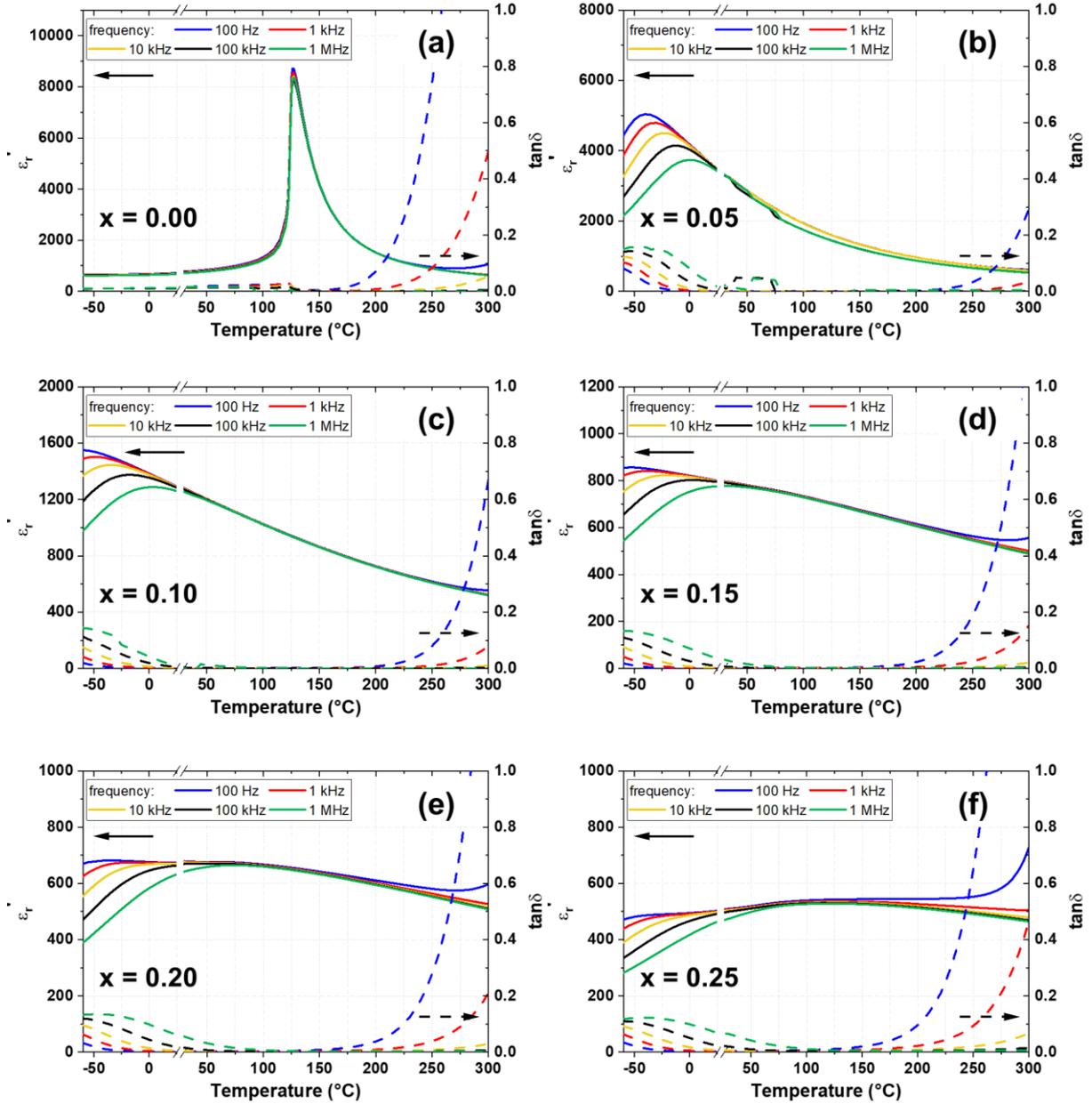

Fig. 5: Temperature-dependent permittivity (solid lines) and loss factors (dashed lines) of BCT-xBMW samples at different frequencies (100 Hz – 1 MHz): (a) x = 0; (b) x = 0.05; (c) x = 0.1; (d) x = 0.15; (e) x = 0.2; (f) x = 0.25

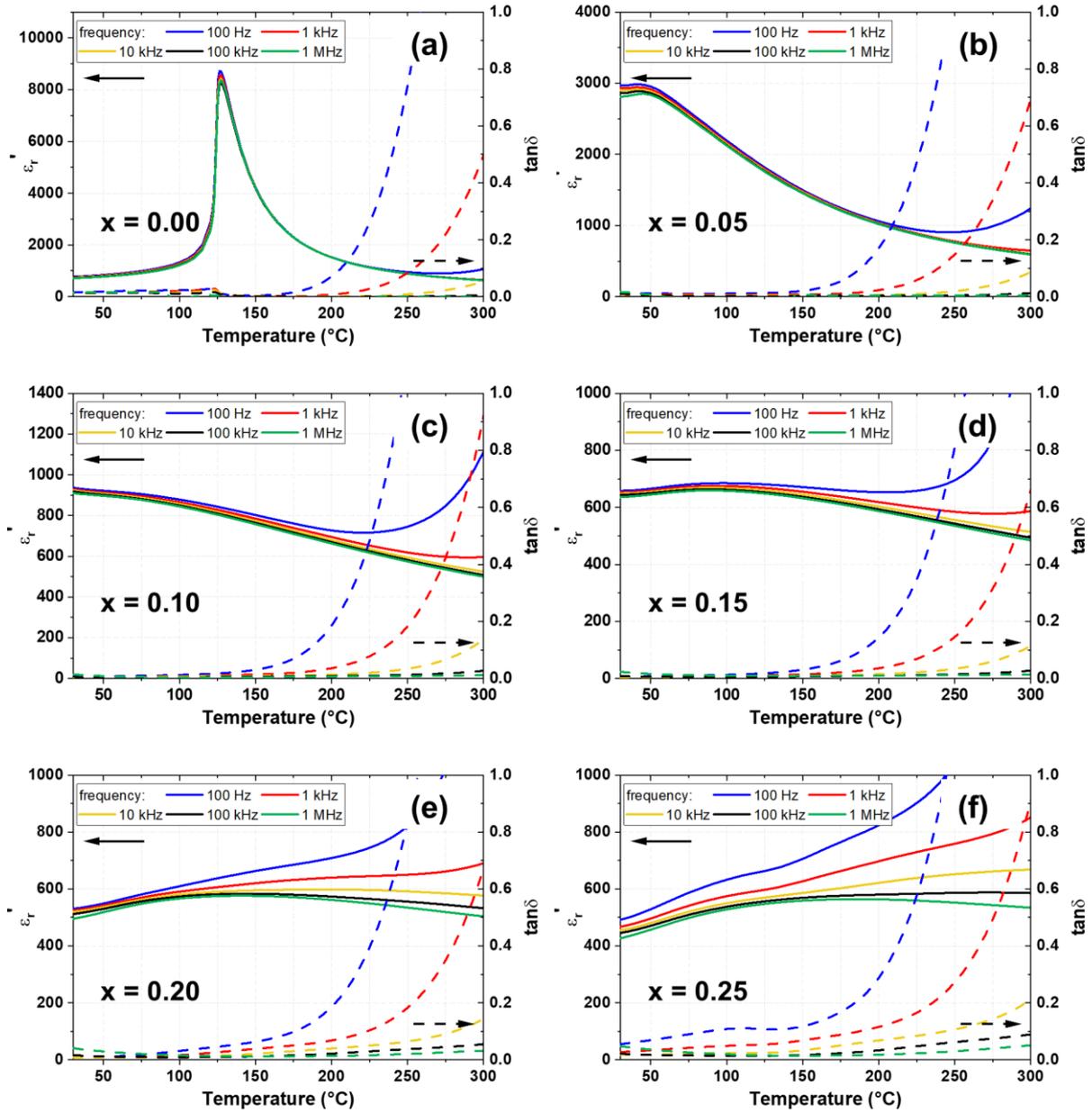

Fig. 6: Temperature-dependent permittivity (solid lines) and loss factors (dashed lines) of BCT-xBMM samples at different frequencies (100 Hz – 1 MHz): (a) x = 0; (b) x = 0.05; (c) x = 0.1; (d) x = 0.15; (e) x = 0.2; (f) x = 0.25

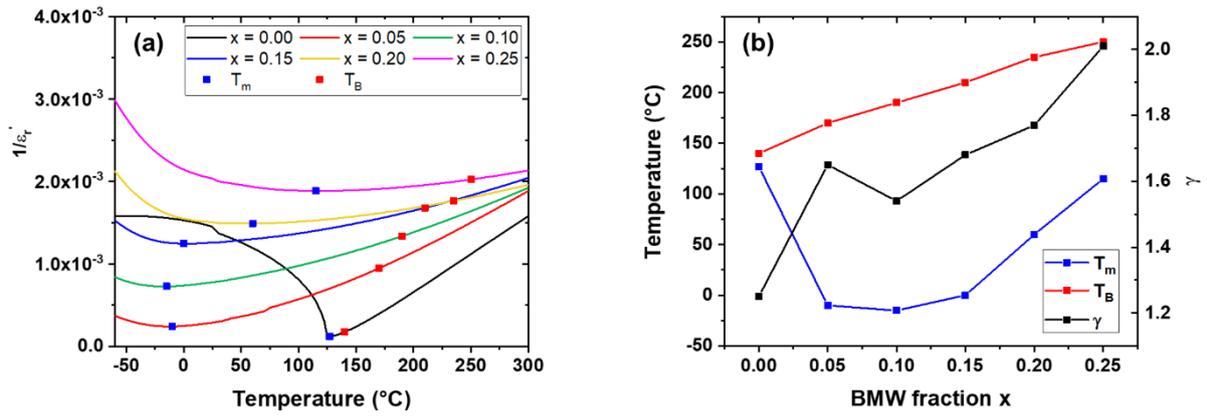

Fig. 7: Curie-Weiss-plot with marked temperature of maximum permittivity ($T_m$) and Burns temperature ($T_B$) (a); trend of $T_m$, $T_B$ and diffuseness exponent γ over compositional variation for BCT-xBMW samples (b) determined from impedance data at 100 kHz

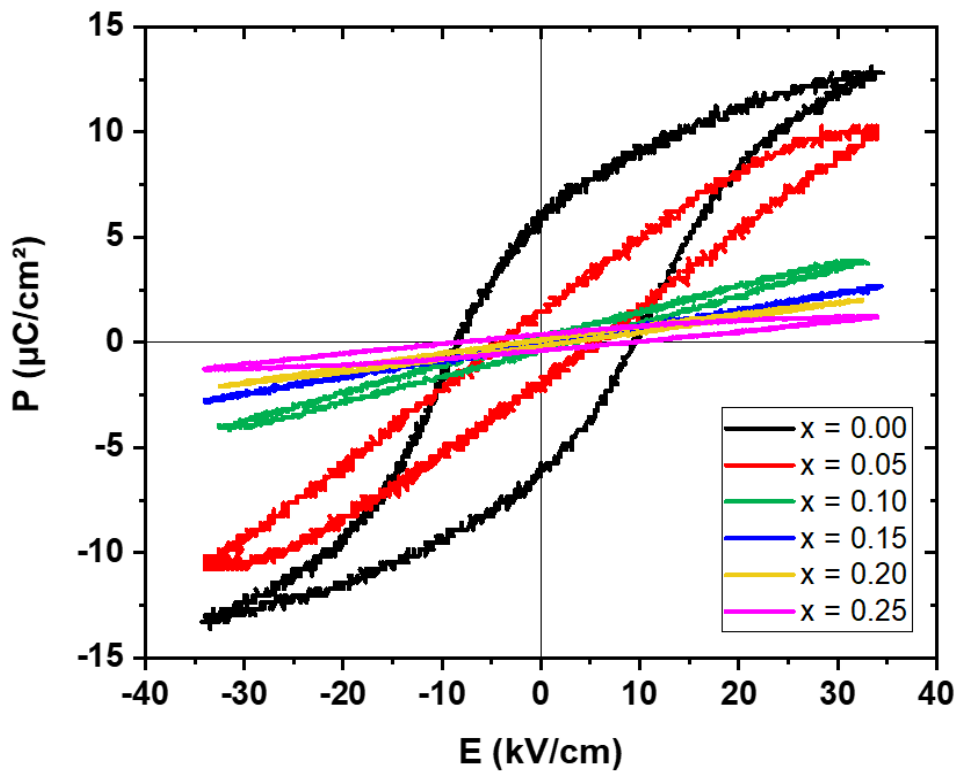

Fig. 8: Polarization vs. field of BCT-xBMW samples at room temperature

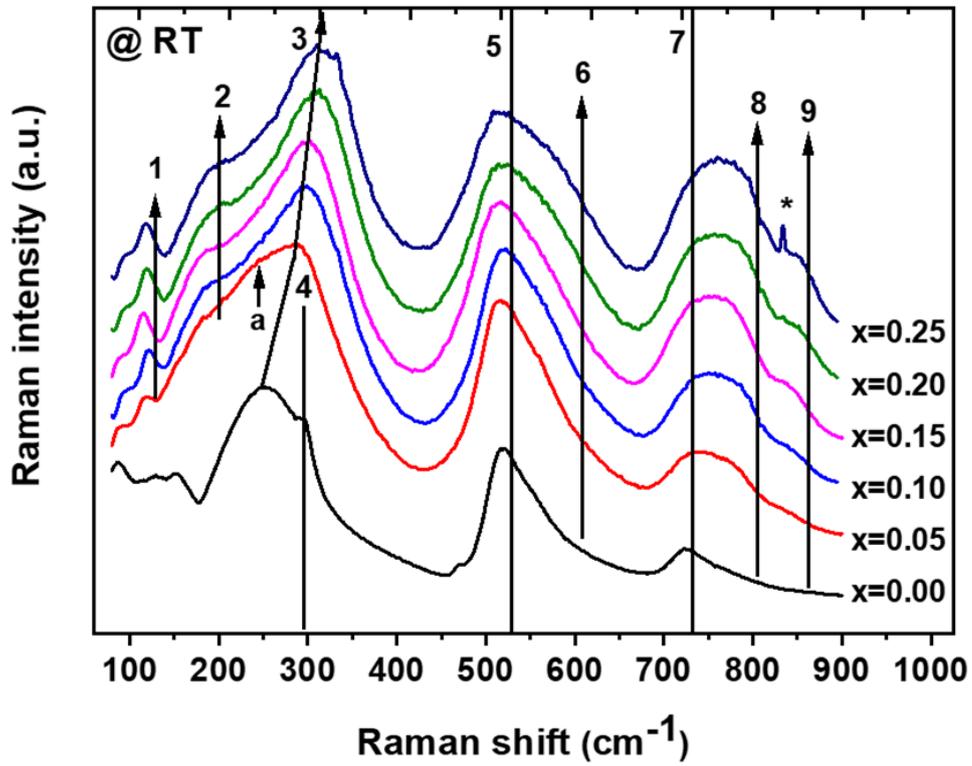

Fig. 9: Room temperature Raman spectra of BCT-xBMW ceramics. The peak marked with an asterisk for the x = 0.25 composition denotes the secondary phase

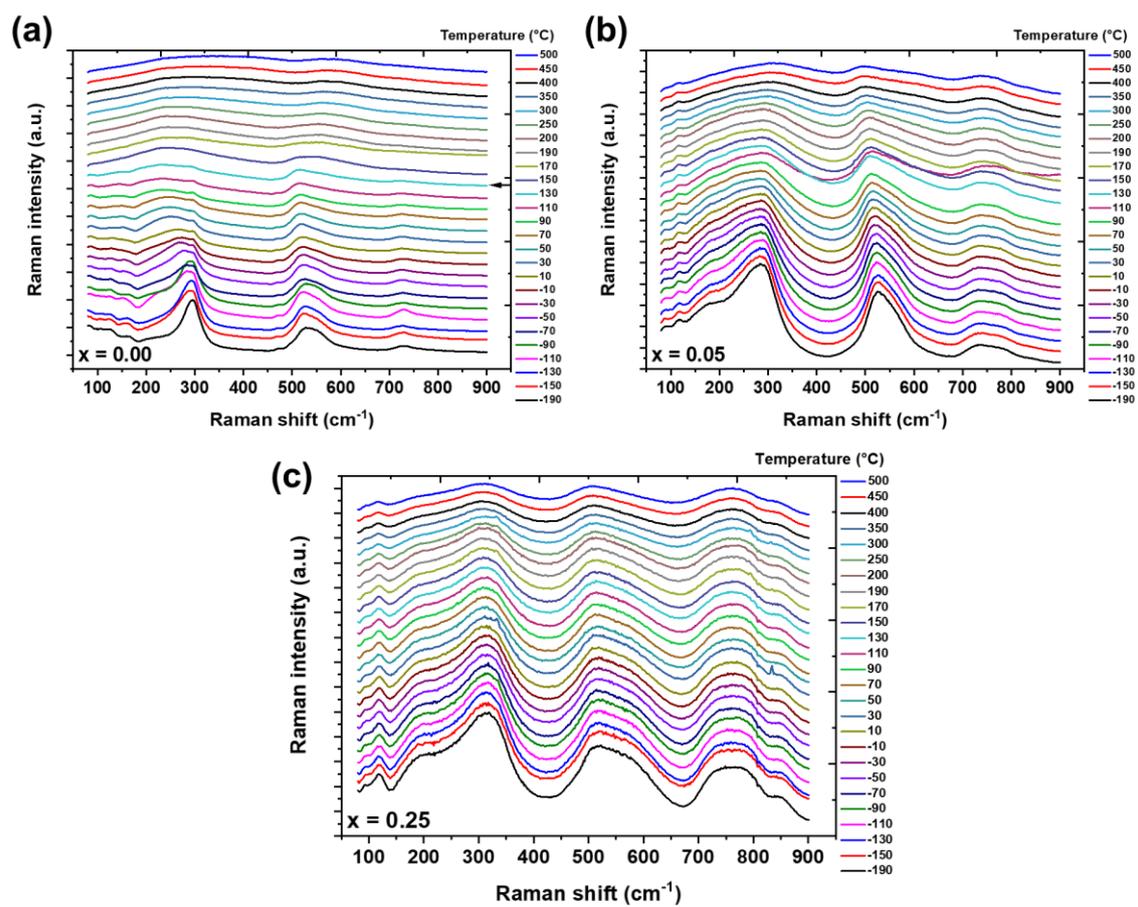

Fig. 10: Temperature dependent Raman spectra for BCT-xBMW: (a) x = 0; (b) x = 0.05 and (c) x = 0.25